\documentclass[aps,prl,preprint,groupedaddress]{revtex4-2}
\usepackage{amsmath}
\usepackage{subcaption}

\usepackage{graphicx}
\usepackage{array}
\graphicspath{ {./figures/}}
\begin{document}

\title{Entropic Analysis to Assess impact of Policies on Disorders and Conflicts within a system: Case Study of Traffic intersection as 12-Qubit Social Quantum System}


\author{Rakesh Kumar Pandey}

\affiliation{Kirori Mal College, University of Delhi, Delhi - 110007}


\date{\today}

\begin{abstract}
Entropic analysis of a scenario at a traffic intersection is attempted in detail. The model is utilized to define \textit {Conflict Entropy}. It is shown that with the use of strategies (policies) like installing \textit{traffic lights} and construction of \textit{flyovers} the Entropy is reduced thereby making the traffic ordered. It is shown that these policies help in reducing the Entropy and eliminating the \textit{Conflict Entropy} completely in both the cases. Such an analysis can find immense application in deciding a favorable policy and in formulation of artificial intelligence algorithms. A striking similarity of the traffic intersection is found with \textit{Quantum systems} of twelve qubits that opens up a new scope of study of traffic flows to understand the behavior of \textit{Quantum systems}.
\end{abstract}


\maketitle

\section{Introduction}
Attempts have been made in the past to study the behavior of society through thermodynamic modeling \cite{Castellano,Loet,Helbing}. Researchers believe that a better and scientific understanding of a society can be attained if such a model can be found \cite{Kenneth,Economics,Perc,Statistical}. It is generally accepted that one must be able to map the parameters of a society with those of a thermodynamic system to expect any meaningful analysis of societal behavior by applying theories of thermodynamics on such a model. Surprisingly, the concept of \textit {Entropy} can potentially find immense application into the mathematical modeling of societal behavior and analysis.\\ 
Entropy is a physical quantity that can be used for making comparison between the extent of '(dis)order' in two different states of the same social system. Any effective policy implemented by a policy enforcing agency would affect the entropy of the system as the status of (dis)order of the social system. A policy, that is supposed to make a system behave systematically and in an ordered manner must be associated with decrease in entropy of the system. On the other hand, a policy that is aimed at enforcing equity or equality must be associated with increase in the entropy of the social system under consideration. Just as we define Entropy in thermodynamics, we can define Entropy of a social system. If there are N possibilities in which a system can stay alternatively with probabilities $P_1, P_2, P_3, ......P_N$ respectively in those states, then the Entropy of the system is given by,
\begin {equation} \label {ref:eq1} 
E= \sum^{i=N}_{i=1} P_i \ ln(P_i)
\end {equation}
A policy for a society targets either to increase or decrease the entropy of an isolated sub-system of a society thereby meaning that it can either create disorder or order in the society. 

\section {Traffic intersection as a 12-Qubit Quantum System}
\begin{figure}[h]
	\includegraphics[width=.5\linewidth]{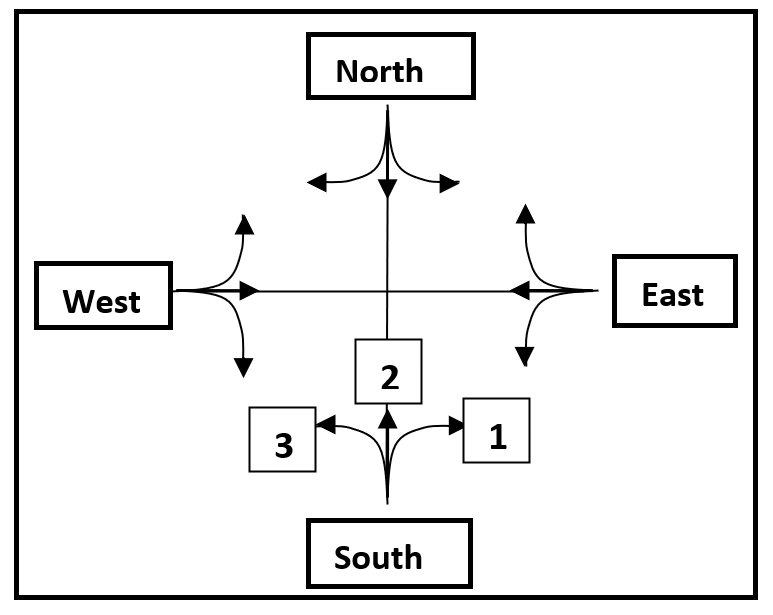}
	\caption {The figure shows twelve possible flows on a traffic intersection}
	\label{fig:directions}
\end{figure}

To analyze and understand policies in terms of its effect on the Entropy of the system, let us consider the behavior of traffic at an intersection as shown in the figure \ref{fig:directions}. In this diagram, there are following four flows of traffics:
\begin{enumerate}
	\item From North towards South (NS flow)
	\item From South towards North (SN flow)
	\item From East towards West and (EW flow)
	\item From West towards East (WE flow)
\end{enumerate}
Each of the above flows has three possibilities:
\begin{enumerate}
	\item Turning Right
	\item Continuing Straight or
	\item Turning Left
\end{enumerate}

These makes the following twelve flow possibilities that will contribute to creating different scenarios at the intersection. The twelve flows can be denoted by NS1, NS2, NS3, SN1, SN2, SN3, EW1, EW2, EW3, WE1, WE2, WE3 with obvious meaning. Since these flows can either be present or absent, it can be denoted by 1 or 0 respectively. Let the probability of a state of the traffic intersection be represented by
\begin{widetext}
	\begin{equation}
	\begin {split}
	P_
	{NS1,NS2,NS3,SN1,SN2,SN3,EW1,EW2,EW3,WE1,WE2,WE3}= \\ P_{NS1} P_{NS2} P_{NS3} P_{SN1} P_{SN2} P_{SN3} P_{EW1} P_{EW2} P_{EW3} P_{WE1} P_{WE2} P_{WE3}
	\end {split}
	\end{equation}
\end{widetext}
This would mean that $P_{000000000000}$ denotes probability of a scenario when all the 12 flows are absent and the intersection is without any traffic and $P_{111111111111}$ would represent probability of the state when traffic from all possible directions will arrive at the intersection simultaneously. Total number of different scenarios at the traffic intersection can thus be determined easily. Since there are 12 flows that can either be present or absent, there will be $2^{12}=4096$ possibilities. But unlike the classical way of being either completely '0' or '1' these would be associated with 'p' and 'q' (with $q=1-p$) probabilities just as it happens in the case of Qubits of Quantum Systems. The traffic intersection, thus also becomes a candidate to analyse and experiment with the behaviour of a 12-Qubit Quantum System. 

\section{Entropic Analysis of Traffic Intersection}
As described above, the twelve flows with different probabilities lead to different scenarios (states) at the traffic intersection. The traffic intersection therefore will always be in one of the $2^{12} = 4096$ states as each of the 12 possibilities may or may not be present (two options). If $P_{abcdefghijkl}$ is chosen to denote the respective probabilities of these 4096 states, the Entropy of the Traffic intersection will be given by the following expression:
\begin {equation}
Enropy = - \sum^1_{a=0}\sum^1_{b=0}\sum^1_{c=0}\dots \sum^1_{l=0}\ P_{abcdefghijkl}\ ln P_{abcdefghijkl}
\end {equation}
Mathematically, it is easy to show that the Entropy of traffic intersection will be maximum when all the 4096 possibilities will have same probability. In this situation the Entropy will be
\begin {equation}
Entropy_{max} = - \sum^1_{a=0}\sum^1_{b=0}\sum^1_{c=0}\dots \sum^1_{l=0}\ (1/4096) ln (2^{-12})
\end {equation}
\begin {equation}
E = -(2^{12}/4096)\ ln(2^{-12}) = 8.317 
\end {equation}

\begin{figure}[ht]
	\begin{subfigure}{.4\textwidth}
		\includegraphics[width=.8\linewidth]{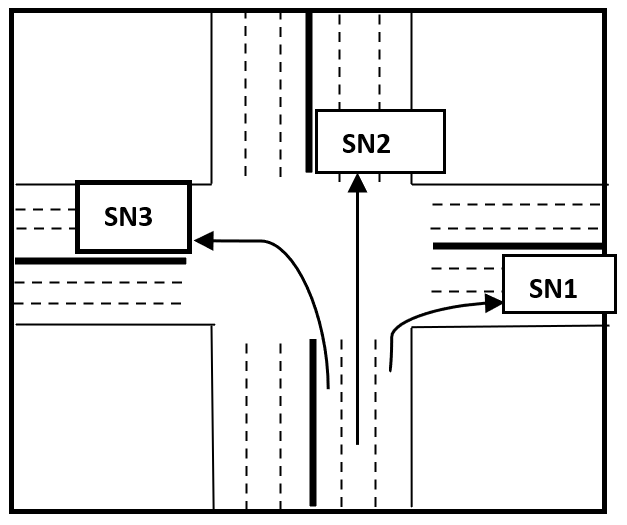}
		\caption {The figure shows lane following traffic flow from south to north}
		\label{fig:phase_1}
	\end{subfigure}
	\begin{subfigure}{.4\textwidth}
		\includegraphics[width=.8\linewidth]{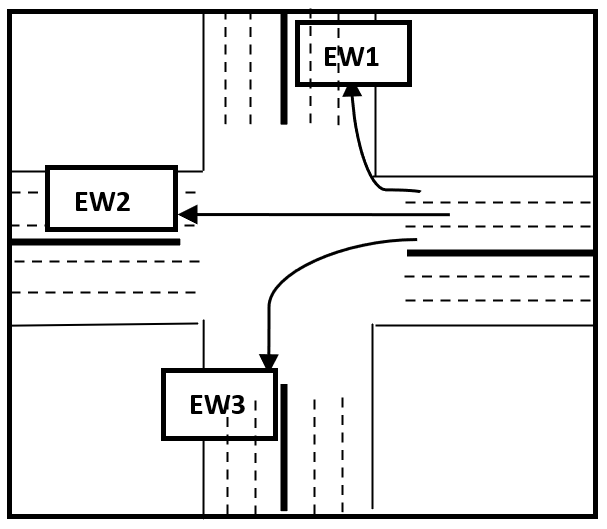}
		\caption {The figure shows lane following traffic flow from east to west}
		\label{fig:phase_3}
	\end{subfigure}
	\begin{subfigure}{.4\textwidth}	
		\includegraphics[width=.8\linewidth]{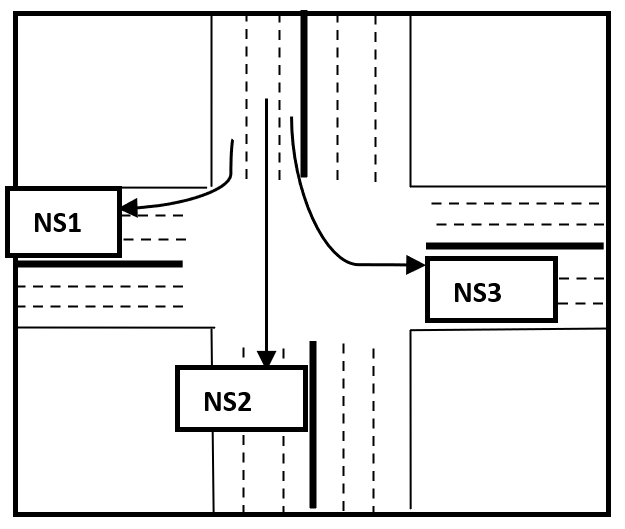}
		\caption {The figure shows lane following traffic flow from north to south}
		\label{fig:phase_2}
	\end{subfigure}
	\begin{subfigure}{.4\textwidth}
		\includegraphics[width=.8\linewidth]{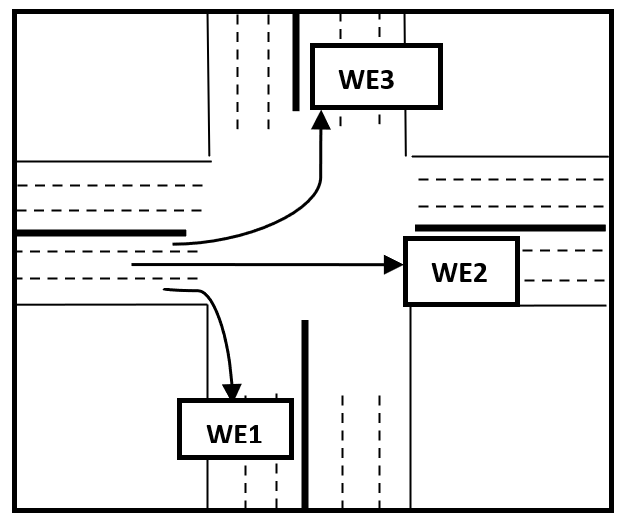}
		\caption {The figure shows lane following traffic flow from west to east}
		\label{fig:phase_4}
	\end{subfigure}
	\caption{lane following traffic}
\end{figure}

\begin{figure}[ht]
	\begin{subfigure}{.4\textwidth}
		\includegraphics[width=1.0\linewidth]{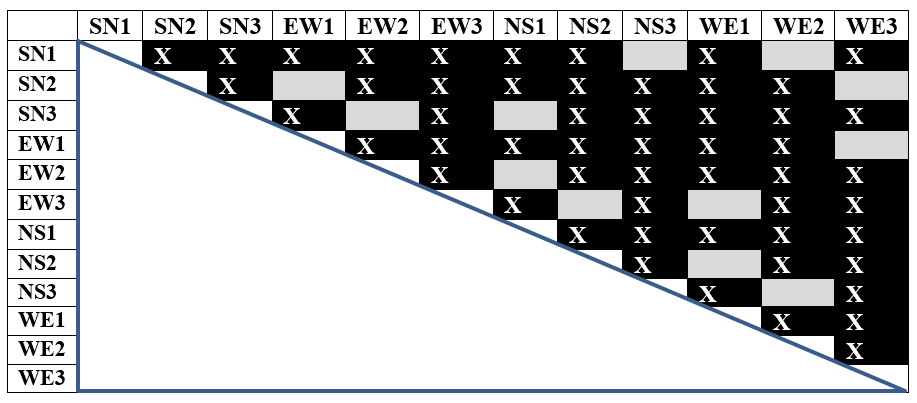}
		\caption {Possible Conflicts in lane-less traffic flow represented by 54 dark boxes out of the 66.}
		\label{fig:conflict_table_2}
	\end{subfigure}
	\begin{subfigure}{.4\textwidth}
			\includegraphics[width=1.0\linewidth]{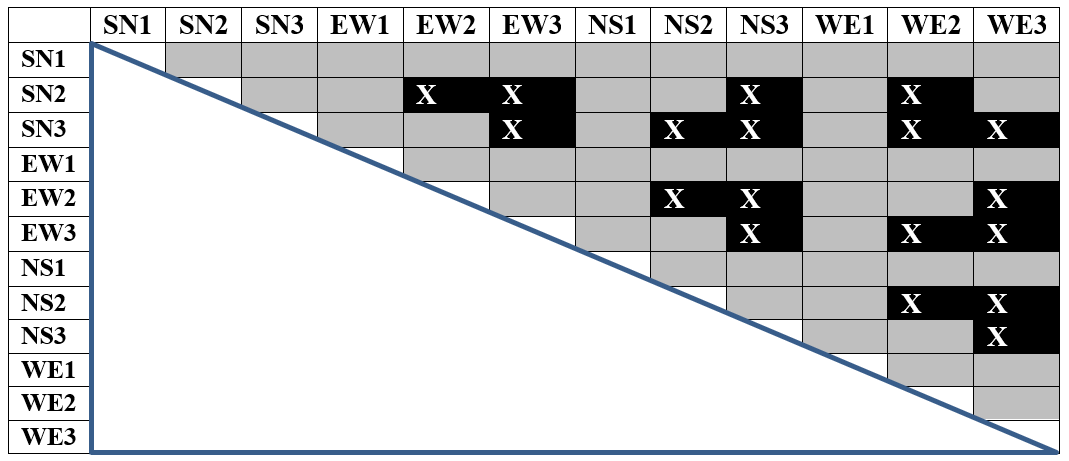}
			\caption {Possible Conflicts in lane following traffic flow represented by 18 darkened boxes out of total 66 of them. }
			\label{fig:conflict_table}
	\end{subfigure}
			\caption {Only upper triagonal is shown of these otherwise symmetric matrices}
\end{figure}

Here it is interesting to notice that there are some pair of flows wherein the two flows will lead to collision between them, if they occur simultaneously. Such pairs define \textit {conflicts} at the intersection as these conflicts can potentially stop the flow. In case of lane-less traffic flow, each of the 12 flows would be in conflict to all others except itself. In the present analysis, 'soft conflicts' have been ignored that occur between two flows in the same direction when overtaking is attempted. If the traffic flows are following lane-driving as shown in the Fig  \ref{fig:phase_1},\ref{fig:phase_2},\ref{fig:phase_3},\ref{fig:phase_4}, then for example, while NS1 will not be in \textit {conflict} with any other traffic flow, NS2 will be in direct \textit {conflict} with the flows represented by EW2, EW3, NS3 and WE2. 
Conflict table both for the lane-less as well as the lane-following traffic flows are shown in the Figure \ref{fig:conflict_table_2}, \ref{fig:conflict_table}. The table is made by identifying all possible conflict combinations and denoting them by blackened boxes with a white cross \textbf {(X)}. Figue \ref{fig:directions} displays the case of lane-less traffic using which the conflict matrix as displayed in the figure \ref{fig:conflict_table_2} is drawn. It is clear that while 54 out of 66 combinations that are possible among 12 possibilities ($^{12}C_2 = 66$) will be in a state of conflict for lane-less traffic the lane-following traffic flow however will have only 18 conflicting pairs. In these analysis conflicts arising due to \textit {overtaking traffic} have not been considered as \textit {conflicts} since flows can continue even if overtaking is not allowed.
		Using this figure one can further determine the number of states of the traffic intersection in which at least one of these conflicts will be present in both the cases. For lane-less traffic this turns out to be 4066 out of 4096 and for lane-following traffic it is 3712 out of 4096 (see Appendix A). This makes the traffic intersection in 99.27 percent conflicting state in case of lane-less traffic. The lane-following traffic also leaves the intersection in conflicting state for 90.63\% of the cases. Although for lane-following traffic, 18 out of 66 combinations appear to be computing as only about 27 percent conflict but it actually becomes 90.63 percent as it creates 3712 conflicting scenario out of a total 4096.
		Using the above analysis although both lane-less and lane-following traffic will have same Entropy, it is the conflict entropy that changes in the two conditions. And therefore, the \textit {\textbf{Conflict Entropy}} can be determined as
		\begin {equation} \label {lane less entropy}
		E_{conflict} (lane_{less}traffic) = \sum^{4096}_{\alpha = 1} P_\alpha ln P_\alpha
		\end {equation}
		\begin {equation} \label {lane following enropy}
		E_{conflict} (lane_{following}traffic) = \sum^{3712}_{\alpha = 1} P_\alpha ln P_\alpha
		\end {equation}
\begin{figure}
	\includegraphics[width=.8\linewidth]{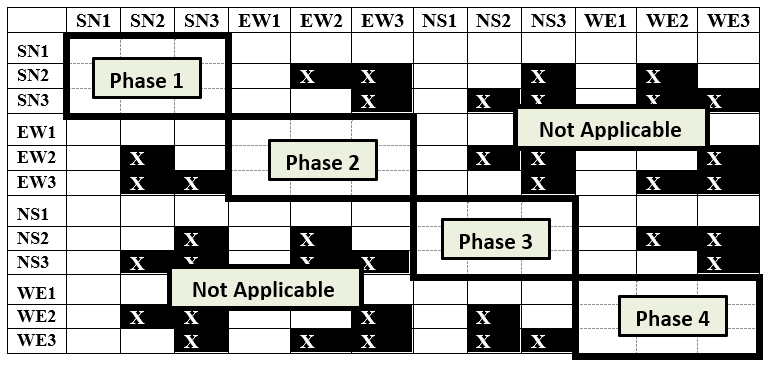}
	\caption {After installing Traffic lights the four phases as shown in the figure (with four 3x3 matrices) will remain valid rendering all other boxes representing invalid scenarios}
	\label{fig:traffic_light}
\end{figure}
		
		Where $\alpha$ takes on all the combinations of \textit {abcdefghijkl} for which \textit {conflict} would take place at the intersection. The non-conflicting Entropy $E_{Non-Conflict}$ will therefore be given by $Entropy - E_{Conflict}$. This will be zero for lane-less traffic but non-zero for lane following traffic.
These \textit {conflicts} get minimised when the occurrence of those 3712 possibilities at the intersection gets reduced by implementing strategies (policies). For example, if one stops East West flow of traffic by barricading the traffic, then we will be left with only two options such as NS2 and SN2. These would be in conflict to each other in the case of lane-less driving but for lane following driving these two will not be in conflicting mode. But this strategy would stop all other flows which would hardly be desirable as the traffic will get accumulated making this policy unsustainable.
However, without completely stopping any of the flows, the \textit {conflicts} can be resolved through two different policies:
\begin{itemize}
	\item Use of Traffic signals:\\
	This is a strategy wherein traffic flows are regulated as shown in the Fig \ref{fig:traffic_light} by making them spread time-wise so that at a given time only non-conflicting traffic-flows are allowed. This is efficiently achieved by implementing traffic signals (lights). Signals allow non-conflicting flows one after the other and permits all the flows in one cycle consisting normally of four phases. In terms of Physics, this means that the system is made to split in four isolated sub-systems such that none of them contains conflicting flows. The conflict table shown in the figure \ref{fig:traffic_light} offer such an idea wherein the flows get time-wise selected using traffic signals/lights and all the flows are allowed to flow for some fixed time exclusively in one of the four phases of this cycle. The cycle keeps on repeating to maintain the arrangement always non-conflicting and at the same time, sustainable.
	
	The Entropy analysis of such an arrangement can be done as follows:
	Since the system (intersection) is time-wise spread into four sub-systems and in each of them only three flows are permitted. Let in the first phase, only NS1, NS2 \& NS3 are allowed. After that in the second phase, WE1, WE2 \& WE3 are allowed and after that SN1, SN2 \& SN3 followed by EW1, EW2 \& EW3. In each of these four phases the entropy of the intersection will be given by:
	\begin {equation}
	Entropy = - \sum^1_{a=0}\sum^1_{b=0}\sum^1_{c=0}\dots \sum^1_{l=0} P_{abc}\ ln (P_{abc})
	\end {equation}
	Where abc will take relevant notations in the four phases as described earlier. Maximum Entropy can be determined when all the probabilities are same. And that will be given by $3 ln (2) = 2.079$. 
	
	\item	By constructing flyovers:\\
	\begin{figure}[h]
		\includegraphics[width=.5\linewidth]{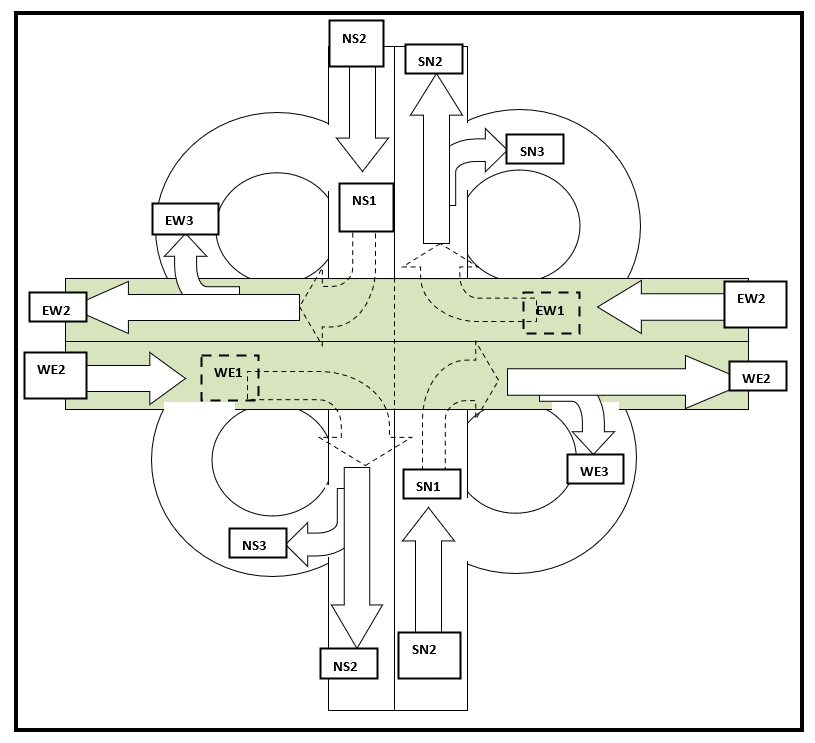}
		\caption {The figure displays the infrastructural expansion that is needed to install flyovers and to allow all the twelve flows to occur simultaneously}
		\label{fig:flyover}
	\end{figure}
	
	In this policy the system gets split into four subsystems but in this case the split is associated with space-wise expansion as shown in the figure \ref{fig:flyover}. This strategy (policy) would allow the traffic to flow simultaneously without ‘conflicts’. This is achieved by constructing a flyover infrastructure. 
	
	\begin{center}
		\begin{table}[!htbp]
			\begin{tabular}{ | m{2.2cm} | m{2.2cm} | m{2.2cm} | m{2.2cm} | m{2.2cm} | m{2.2cm} | m{2.2cm} | }
				\hline
				Traffic Flow & Capacity to handle vehicles/min & Sample A: actual no. of vehicles per min & Probability (p) of traffic in Sample A & Sample B (no. of vehicles) & Sample B (phase 1) with traffic lights & Sample B after flyover construction \\ 
				\hline
				NS1	& 50 & 25 & 0.5 & 8 & 32 & 8 \\ 
				\hline
				NS2 & 60 & 30 & 0.5 & 12 & 48 & 12 \\ 
				\hline
				NS3 & 40 & 20 & 0.5 & 7 & 28 & 7 \\ 
				\hline
				EW1 & 50 & 25 & 0.5 & 9 & 0 &  \\ 
				\hline
				EW2 & 60 & 30 & 0.5 & 14 & 0 &  \\ 
				\hline
				EW3 & 40 & 20 & 0.5 & 8 & 0 &  \\ 
				\hline
				SN1 & 50 & 25 & 0.5 & 5 & 0 &  \\  
				\hline
				SN2 & 60 & 30 & 0.5 & 13 & 0 &  \\  
				\hline
				SN3 & 40 & 20 & 0.5 & 7 & 0 &  \\  
				\hline
				WE1 & 50 & 25 & 0.5 & 10 & 0 &  \\  
				\hline
				WE2 & 60 & 30 & 0.5 & 14 & 0 &  \\  
				\hline
				WE3 & 40 & 20 & 0.5 & 9 & 0 &  \\  
				\hline
				\multicolumn{3}{|c|}{Total Entropy} & 8.317 & 5.807 & 1.765 & 1.404 \\
				\hline
				\multicolumn{3}{|c|}{Conflict Entropy for lane-less traffic} & 8.293 & 5.481 & 0 & 0 \\
				\hline
				\multicolumn{3}{|c|}{Conflict Entropy for lane-following traffic} & 7.538 & 2.860 & 0 & 0 \\
				\hline 
			\end{tabular}
			\caption{ Entries in the columns of this table are as follows. 
				Column 1: Traffic Flow directions, Column 2: Capacity of the intersection to serve vehicles per minute in that lane, Column 3: Number of vehicles for maximum Entropy in Sample A, Column 4: Probability p of Sample A, Column 5: Actual number of vehicles in Sample B, Column 6: Number of vehicles in the Phase 1 when traffic signals were installed for Sample B, Column 7: Number of vehicles for the traffic flowing in the space specified for the traffic flowing from North to South after flyover construction}
			\label{Table one}
		\end{table}
	\end{center}
	The intersection itself gets effectively split into four intersections each having Maximum-Entropy again given by 2.079 as explained above. A lower Entropy means an ordered system. Total Maximum-Entropy of the system can be obtained by adding Entropies of all the four sub-systems and that is still 8.317.
\end{itemize}
In conclusion to the entire discussion Table \ref{Table one} is shown displaying quantitative analysis of a Traffic intersection. This table is self explanatory and shows how the intersection can be analysed for two samples A \& B. Sample A represents the state with highest Entropy (when all the states have equal probability). Sample B is a realistic sample that is analysed on the basis of Entropy (using equaion \ref{ref:eq1}) and Conflict Entropy both under lane-less and lane-following traffic conditions (using equations \ref{lane less entropy} and \ref{lane following enropy}) when traffic lights are installed and when flyover gets constructed respevtively. As the Sample A is associated with maximum entropy, any drift towards equal probabilities of the states, as a result of some policy will get reflected in increase of the entropy and will eventually make the system more disordered.

\section{scopes and limitations: discussion}
It will be interesting to analyse real cases in real time wherein all the probabilities will not usually be same and will get distributed arbitrarily in all the twelve flows. It would also be interesting to find out the conditions on these probabilities for keeping the traffic signal solution possible.

Such an entropic analysis can also find application in analysing and comparing policies and their performances. A decision in favour of implementing a policy can be based on its expected impact on the Total Entropy and Conflict Entropy of the system in consideration.

The calculation of entropy as given by the equation \ref{ref:eq1} involves exponential time complexity. It needs to do $2^{12}$ additions to calculate Entropy of the 12-Qubit system represented by Traffic Intersection. This time complexity will limit implementation of such an analysis for large systems until Quantum Computers are reralized.

Interestingly, the traffic intersection offers a good candidate for a quantum state with 12 ‘Qubits’. Between two crossings the quantum state of the traffic will remain coherent if ttraffic is restricted from entering using fences on the two sides. Study of the behaviour of traffic at an intersection may probably allow us to develop a better understanding of quantum systems. 
\subsection{}
\subsubsection{}

\section{acknowledgement}
\begin{acknowledgments}
	I thank Kirori Mal College to have provided me the required environment to carry out this work.
\end{acknowledgments}


\section{Appendix}
{Number of states with ‘conflicts’ at the intersection in which SN2 is necessarily present is determined from observing the conflict table that shows four flows against SN2 and they happen to be EW2, EW3, NS3 \& WE2. Total number of such possibilities will be given by picking up at least one out of these four given by $^4C_1 +^4C_2 +^4C_3 +^4C_4 =2^4-1$. And for each one of these the rest seven flows combine in $2^7$possible ways. This makes a total of $(2^4-1)2^7$ possible scenarios. Similarly, to determine the number of conflicting states having SN3 necessarily at the intersection will have $(2^5-1)2^5$. While there are five conflicting flows for SN3 now we must exclude SN2 from rest of the flows since those have already been counted above. Proceeding in the same way, conflicting possibilities with EW2 in them will be given by $(2^3-1)2^6$, with EW3 in them as $(2^4-1)2^7$, with NS2 in them as $(2^4-1)2^7$ \& with NS3 necessarily included as $2^5$. Total conflicting scenarios will thus be counted as $(2^4-1) 2^7+(2^5-1) 2^5+(2^3-1) 2^6+(2^3-1) 2^5+(2^2-1) 2^5+2^5=3712$.\\

For lane-less traffic similar calculations would estimate $(2^9-1) 2^2+(2^8-1) 2^2+(2^7-1) 2^2+(2^7-1) 2+(2^6-1) 2+(2^4-1) 2^2+(2^5-1) +(2^3-1) 2+(2^2-1) 2+(2^2-1) =4066$ conflicts.}


\end{document}